\documentclass[a4paper,
preprint,
preprintnumbers,
nofootinbib,
 amsmath,amssymb,
]{revtex4-1}

\usepackage{cancel}
\usepackage{graphicx}
\usepackage{tabularx}
\usepackage{colortbl}
\usepackage{dcolumn}
\usepackage{bm}
\usepackage{latexsym}
\usepackage{amsmath,amsfonts,amssymb}
\usepackage{graphicx,epsfig}
\usepackage{subfig}
\usepackage{psfrag}
\usepackage[normalem]{ulem}
\usepackage{hyperref}
\usepackage{slashed}
\usepackage[mathlines]{lineno}
\usepackage{tikz} 
\usepackage{tikz-feynman}
\tikzfeynmanset{compat=1.1.0}
\usepackage{color}
\usepackage{braket}
\usepackage[english]{babel}
\usepackage{array}

\definecolor{dgreen}{rgb}{0,0.39,0}

\begin{document}
\title{Quantum field theory with the generalized uncertainty principle II: Quantum Electrodynamics}

\author{Pasquale Bosso}
\email{pasquale.bosso@uleth.ca}
\affiliation{Theoretical Physics Group and Quantum Alberta,
Department of Physics and Astronomy,
University of Lethbridge,
4401 University Drive, Lethbridge,
Alberta, T1K 3M4, Canada}

\author{Saurya Das}
 \email{saurya.das@uleth.ca}
\affiliation{Theoretical Physics Group and Quantum Alberta,
Department of Physics and Astronomy,
University of Lethbridge,
4401 University Drive, Lethbridge,
Alberta, T1K 3M4, Canada}

\author{Vasil Todorinov}
 \email{v.todorinov@uleth.ca}
\affiliation{Theoretical Physics Group and Quantum Alberta,
Department of Physics and Astronomy,
University of Lethbridge,
4401 University Drive, Lethbridge,
Alberta, T1K 3M4, Canada}

 
\date{\today}

%
%
%
%

\begin{abstract}
 Continuing our earlier work on the 
 application of the Relativistic Generalized Uncertainty Principle (RGUP) to quantum field theories, in this paper we study 
 Quantum Electrodynamics (QED) with minimum length.
We obtain expressions for the Lagrangian, Feynman rules and scattering amplitudes of the theory, and discuss their consequences for 
current and future high energy physics experiments.   
We hope this will provide an improved 
window for testing Quantum Gravity effects in the laboratory.

\end{abstract}
\pacs{Valid PACS appear here}
%
\maketitle

\newpage

\section{Introduction}

\noindent 

The existence of a minimum length has been predicted by \emph{gedanken} experiments in black hole physics and almost all theories of Quantum Gravity (QG), such as String theory (ST) and Loop Quantum Gravity (LQG)
\cite{Physics1988,Magueijo:2004vv,Rovelli:1994ge,Rovelli:1990wq,Ashtekar:1991mz,Amelino-Camelia2010,Smolin2003,Rovelli:1997yv,Amati:1988tn,Maggiore:1993kv,Maggiore:1993rv,Witten:2001ib,AHARONY2000183}.
In fact, it is considered to be one of the most robust predictions of QG theories.
Phenomenological implications of theories with minimum measurable length are often studied via a modification of the Heisenberg uncertainty relation to accommodate a minimum uncertainty in position and a minimum length.  
This modification, known as Generalized Uncertainty Principle (GUP), has been studied in a series of papers, among which \cite{Kempf1995,Ali2011,Bosso2018,Ali2010,Das2011,Das2008,Das2014,BossoPhD}.
However, apart from some exceptions  \cite{Hossenfelder2006,Deriglazov2014,Kober:2010sj,Kober:2011dn,Shibusa:2007ju,Husain:2012im,Faizal:2017map,Quesne2006},
most of the effort has been directed to non-relativistic models.
Other notable approaches to the problem of a minimum position uncertainty from other perspectives are in \cite{Zakrzewski1994,Pramanik2013_1,Pramanik2014_1,Bosso2018_2,Capozziello:1999wx,Snyder:1946qz,Szabo:2006wx,Chaichian:2004za}.
Nonetheless, theories with non-relativistic GUP are affected by a frame-dependent minimum length, effectively bringing back the concept of aether.
The frame-independence of a minimum length is achieved through a relativistic extension of GUP as follows 
\begin{equation}
\label{GUPxp}[x^{\mu},p^{\nu}]=i\hbar\,\left(1+\gamma  p^{\rho}p_{\rho}\right)\eta^{\mu\nu}+i\,\hbar\gamma p^{\mu}p^{\nu}\,,
\end{equation}
where $\gamma=\frac{\gamma_0}{(M_{\mathrm{Pl}}\,c)^2}$, $M_{\text{Pl}}$ is the Planck mass and $\gamma_0$ is a dimensionless parameter used to set the scale for the quantum gravitational effects. This is the   most general form of quadratic  Relativistic Generalized Uncertainty Principle (RGUP), which is used in this work. In natural units, \emph{i.e.} $\hbar=c=1$, one has $\gamma =\gamma_0\, \ell_{\mathrm{Pl}}^2$, where  $\ell_{\mathrm{Pl}}$ is the Planck length.
As typical in GUP, it can be observed that the momentum and position are no longer canonically conjugate.
Therefore, one can introduce the canonically conjugate auxiliary variables $x_0^\mu$ and $p_0^\mu$, such that
\begin{equation}
\label{MathVar}p_0^{\mu} = -i\frac{\partial}{\partial x_{0\,\mu}}, \quad
[x_0^{\mu},p_0^{\nu}] = i\eta^{\mu\nu}\,.
\end{equation}

In a previous work 
\cite{Todorinov:2018arx}, 
the authors provided a Lorentz invariant minimum length.
Furthermore, they obtained some interesting properties of spacetime, such as its non-commutativity at high energies.
Moreover, a relationship between the parameters of the RGUP was found, allowing the non-commutative spacetime to preserve the symmetries of classical spacetime. In fact  one can easily show that the squared momentum $p_\mu p^\mu$ is still a Casimir invariant of the modified Poincare group.  
Thus, the dispersion relation for the modified Poincar\'e group is preserved as well and has the following form
\begin{equation} \label{DR}
    p^{\rho}p_{\rho}=-(mc)^2\,.
\end{equation}
This result opens the door to the formulation of quantum field theory (QFT) with a minimum length.
In fact, the authors recently presented QFT for complex scalar fields and derived the corresponding Feynman rules in \cite{Bosso2020:AoP}.
An interesting consequence of considering the scalar QFT with a minimum length was the fact that up to six particle vertices were allowed in the theory.
Corrections to the invariant amplitude and the differential cross section of the scattering of a scalar electron and a muon, were found.

In the present work, the authors continue our program by formulating Quantum Electrodynamics (QED) with a minimum length.
This paper is structured as follows:
in the the second section, the Ostrogradsky method is used for the derivation of the QED Lagrangian, as well as the gauge field Lagrangian.
In section \ref{sec:Feynmanrules}, the Feynman rules is derived from the minimally coupled to the gauge field QED Lagrangian.
The transition amplitudes for an three particle vertices are calculated in section \ref{sec:scatteringcrosssecton}. Additionally, calculations for the invariant amplitude for an electron and muon scattering are presented.
This allows for computation of the differential and total cross sections for ultra-relativistic regime.

\section{\label{sec:QED}Quantum Electrodynamics Lagrangian}

Starting from the dispersion relation Eq.\eqref{DR} and  expressing it in terms of the auxiliary variables defined in Eq.\eqref{MathVar}, one gets
 \begin{equation}
    \label{ModKG}
    p_0^{\rho}p_{0\rho}(1+2\alpha\tau^2p_0^{\sigma}p_{0\sigma})=-(mc)^2 \,.
\end{equation}
As derived in the appendix \ref{app:somutions}, one can then prove that the Dirac equation has the form 
 \begin{equation}\label{EoM}
    \left(\tau^\mu p_\mu-m\right)\psi =\left[\tau^{\mu}p_{0\,\mu}(1+\gamma p_{0\rho}p_0^\rho)-m\right]\psi=0\,,
 \end{equation}
 where $\tau^\mu$ are the Dirac matrices and $\psi$ is a Dirac spinor.
The algebra of the Dirac matrices remains unchanged due to the fact that the dispersion relation is unchanged, which allows us to use their usual representation
 \begin{align}\label{gamma}
\tau^0=\begin{pmatrix} 
\mathbf{I}&0 \\
0 & -\mathbf{I} 
\end{pmatrix},\,\,\,\,\,\tau^i=\begin{pmatrix} 
0&\sigma^i \\
\sigma^i &0
\end{pmatrix}\,,
\end{align}
where $\sigma^i$ are the Pauli matrices and $i\in\{1,2,3\}$. More detailed explanations of the calculations are presented in appendix \ref{app:somutions}.
Eq.\eqref{EoM} and its Dirac conjugate
are the equations of motion for the RGUP-QED Lagrangian. One can use the Ostrogradsky method to derive the 
corresponding higher derivative Lagrangian
\cite{deUrries:1998obu,Woodard:2015zca,Pons:1988tj}
as shown in Appendix \ref{app:lagrangian}, obtaining
 \begin{equation}\label{DirackLagrangian}
     \mathcal{L}_{\psi}=\bar\psi\left[ i\tau^\mu\partial_\mu(1-\gamma\,\partial_\rho\partial^\rho) -m\right]\psi\,.
 \end{equation}
Thus, the gauge field Lagrangian has the following form
 \begin{equation}\label{GaugeFieldL}
     \mathcal{L}_{A}=-\frac{1}{4}F^{\mu\nu}F_{\mu\nu}=-\frac{1}{4}F^{\mu\nu}_0F_{\mu\nu 0}-\frac{\gamma }{2}F_{\mu\nu 0} \partial_\rho \partial^\rho F^{\mu\nu}_0
     \,.
 \end{equation}
It is worth noticing that this is gauge invariant with a $U(1)$ gauge group.

\section{\label{sec:Feynmanrules}Feynman rules}

The modified Feynman propagator is the Green's function of the modified Dirac differential operator.
Therefore, from the modified Dirac equation in Eq.\eqref{EoM}, one has\\
 \begin{equation}\label{GreenFunc}
   \left[ i\tau^\mu\partial_\mu(1-\gamma\partial_\rho\partial^\rho) -m\right] G(x-x') = - i \delta(x-x')\,.
\end{equation}
Expressing the Green's function $G(x - x')$ in terms of its Fourier transform
\begin{equation}\label{GreenFourier}
    G(x-x')=\int \frac{d^4p_0}{(2\pi)^4} \tilde G(p_0) e^{-i p_0\cdot(x-x')},
\end{equation}
and substituting it in Eq.\eqref{GreenFunc}, one gets
\begin{equation}
\int \frac{d^4p_0}{(2\pi)^4} \tilde G(p_0)\left[ i\tau^\mu\partial_\mu(1-\gamma\partial_\rho\partial^\rho) -m\right] e^{-i p_0\cdot(x-x')}=-i \int\frac{d^4p_0}{(2\pi)^4}e^{-i p_0\cdot(x-x')}\,.
\end{equation}
Therefore, the Fourier transform of the Feynman propagator has the form
\begin{equation}
     \tilde G(p_0) \left[\tau^{\mu}p_{0\,\mu}(1+\gamma p_{0\rho}p_0^\rho)-m\right] =-i \mathbb{I}^4\,.
\end{equation}
Multiplying both sides by $\left[\tau^{\mu}p_{0\,\mu}(1-\gamma p_{0\rho}p_0^\rho)+m\right]$, one obtains the propagator
%
%
\begin{equation}
     G(x-x')=\int \frac{d^4p_0}{(2\pi)^4}\frac{-i \left[\tau^{\mu}p_{0\,\mu}(1+\gamma p_{0\rho}p_0^\rho)+m\right]}{p_0^{\mu}p_{0\,\mu}(1+\gamma p_{0\rho}p_0^\rho)^2-m^2}e^{-i p_0\cdot(x-x')}\,.
\end{equation}
Further, when the
full RGUP-QED action is considered, it reads as follows 
\begin{align}
   S= \int \mathcal{L}\,d^4x &= \int\left[\mathcal{L}_{A} + \mathcal{L}_{\psi}\right]\,d^4x \nonumber\\ &= \int\bar{\psi}\left[ i\tau^\mu D_{\mu}\psi -\gamma \bar{\psi} i\tau^\mu D_{\mu}D_{\nu}D^{\nu}\psi  - \frac{1}{4}F^{\mu\nu}F_{\mu\nu}\right]\,\,d^4x \,,
   \label{qedaction1}
\end{align}
in which the minimal coupling to the $U(1)$ gauge field has been introduced in the usual form
 \begin{equation}
     \partial_\mu \rightarrow D_\mu=\partial_\mu +ie A_\mu\,.
\end{equation}
It can be easily verified that the above, along with a local phase transformations of $\psi$ and $\bar\psi$, leaves the action
in Eq.\eqref{qedaction1} invariant.
The vertices can be read from the minimally coupled  modified Dirac field Lagrangian Eq.\eqref{DirackLagrangian} 
\begin{multline}\label{RGUPQED}
    \mathcal{L}_{\psi}=i\bar{\psi} \tau^\mu \partial_{\mu}\psi +i\gamma\bar{\psi}\tau^\mu \partial^{\rho}\partial_{\rho}\partial_{\mu}\psi-m\bar{\psi}\psi\\
    -e \left[\bar{\psi} \tau^\mu A_{\mu}\psi - 2 \gamma \bar{\psi} \tau^\mu \left(\partial_{\mu}A^{\rho}\right) \partial_\rho\psi - 2 \gamma\bar{\psi} \tau^\mu A^{\rho}\partial_{\mu}\partial_\rho\psi - \gamma\bar{\psi} \tau^\mu A_\mu\partial^{\rho}\partial_{\rho}\psi\right]\\
    - i e^2 \gamma \left[2 \bar{\psi} \tau^\mu A_\mu A^{\rho}\partial_{\rho}\psi + 2 \bar{\psi} \tau^\mu \left(\partial_\mu A^{\rho}\right)A_{\rho}\psi - \bar{\psi} \tau^\mu A^{\rho}A_{\rho}\partial_\mu\psi\right]\\
    - e^3\gamma\bar{\psi}\tau_{\mu}A^{\mu}A^{\rho}A_{\rho}\psi\,.
\end{multline}
One can see that there are up to five particle vertices.
They include always two fermions and from one to three gauge bosons.
Furthermore, one can see that the coupling constants for each vertex is a product of powers of the electronic charge $e$ and the RGUP coefficient $\gamma$.
In fact, the power of the electronic charge determines how many bosons couple to the vertex and the power of $\gamma$ is $0$ for the usual terms and $1$ for the RGUP corrections terms.
For the purposes of this work, the authors chose to focus on the invariant amplitude and cross section for vertices involving only one gauge boson. 
This includes the 
leading order in the transition amplitudes and RGUP corrections to it.

 \section{\label{sec:scatteringcrosssecton}Electron-muon scattering cross section}
\begin{figure}
\centering
\feynmandiagram [small,baseline=(b.base),vertical=b to a] {
i1 [particle=\(e^{-}\)] -- [fermion] a -- [fermion] i2 [particle=\(e^{-}\)],
a -- [photon, edge label=\(A^\mu \)] b,
f1 [particle=\( \mu^{-} \)] -- [fermion] b -- [fermion] f2 [particle=\( \mu^{-}\)]};
\caption{\label{figure1}The Feynman diagram of the electron, muon  scattering, in $t$ channel. }
\end{figure}
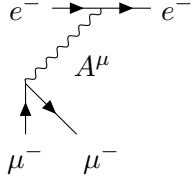

This section presents the calculations of the quantum gravity corrections to the electromagnetic cross section of an electron and a muon, following  the procedure outlined in 
\cite{Halzen:1984mc}. The transition amplitude can be read off
from the three particle interaction terms in Eq.\eqref{RGUPQED}
\begin{equation}
    T_{fi}=-i\int  A^\mu j_\mu\,d^4x\,, \label{amplitude}
\end{equation}
where $j_{\mu}$ is the current corresponding to the electrons and muons. For ease of 
calculations, one splits the transition amplitude for the three particle Feynman vertex into the usual term $ T_{fi}^{(1)}$ and two correction terms  $ T_{fi}^{(2)}$ and $ T_{fi}^{(3)}$ as follows
 \begin{subequations} \label{amplitude_separated}
\begin{align}
 T_{fi}^{(1)}&=i\int e\bar{\psi}_f \tau^\mu A_{\mu}\psi_i\, d^4 x\,,\\
 T_{fi}^{(2)}&=i\int2e\gamma\partial_{\rho}\bar{\psi}_f \tau^\rho A^{\mu}\partial_\mu\psi_i\, d^4 x\,,\\
 T_{fi}^{(3)}&=-i\int e\gamma\bar{\psi} \tau^\mu A_\mu\partial^{\rho}\partial_{\rho}\psi\, d^4 x\,.
\end{align}
\end{subequations}
From Eq.\eqref{amplitude_separated} presented above, one can isolate the current and its corrections terms. They take the following form
\begin{subequations} \label{currentsPsi}
\begin{align}
 j_{fi\,\mu}^{(1)}&=-e\bar{\psi}_f \tau^\mu\psi_i\,,\\
 j_{fi\,\mu}^{(2)}&= -2e\gamma\partial_{\mu}\bar{\psi}_f \tau^\mu \partial_\rho\psi_i\,,\\
 j_{fi\,\mu}^{(3)}&= e\gamma\bar{\psi}_f \tau^\mu \partial^{\rho}\partial_{\rho}\psi_i\,.
\end{align}
 \end{subequations}
In terms of the physical momentum $p_\mu$, the Dirac equation will have the form of Eq.\eqref{EoM}. One can recall that the modified Poincar\'e group has the  squared physical momentum as a Casimir invariant. This means that the  solutions of Eq.\eqref{EoM} need to obey the usual dispersion relation  and will be of the form shown in the appendix \ref{app:somutions}
\begin{equation}
 \psi(x)=u^{(s)}\left(p\right)e^{-ip\cdot x}\,,
 \end{equation}
 where $u^{(s)}\left(p\right)$ are four component complex spinors 
\begin{subequations}\label{freefieldsolutions}
\begin{align}
    u^{(s)}\left(p\right) = & N\begin{pmatrix}
        \chi^{(s)}\\
        \frac{\sigma \cdot p}{E+m}\chi^{(s)}
    \end{pmatrix}\,, & E > & 0\,,\\\nonumber\\
    u^{(s+2)}\left(p\right) = & N\begin{pmatrix}
        \frac{-\sigma \cdot p}{E+m}\chi^{(s+2)}\\
        \chi^{(s+2)}
    \end{pmatrix}\,, & E < & 0\,,
\end{align}
\end{subequations}
where $p$ is the physical four momentum, $E$ is the zeroth component of that vector, $s\in\{1,2\}$, and $\chi^{(s)}$ are 
\begin{equation}
   \chi^{(1)}=\begin{pmatrix}
1\\
0
\end{pmatrix}, \qquad \qquad \chi^{(2)}=\begin{pmatrix}
0\\
1
\end{pmatrix}\,.
\end{equation}
Substituting the form of the field into Eq.\eqref{currentsPsi}, one gets
\begin{subequations} 
\label{currentsU}
\begin{align}
 j_{fi\,\mu}^{(1)}&=-e\bar{u}_f \tau^\mu u_i\,e^{i\left(p_f-p_i\right)\cdot x}\,,\\
 j_{fi\,\mu}^{(2)}&=-2e\gamma\bar{u}_f p_{f\,\rho}\tau^\rho p_{i\,\mu} u_i\,e^{i\left(p_f-p_i\right)\cdot x}\,,\\
 j_{fi\,\mu}^{(3)}&=e\gamma\bar{u}_f \tau^\mu p_{i\rho}p^{\rho}_i u_i\,e^{i\left(p_f-p_i\right)\cdot x}\,.
\end{align}
\end{subequations}
Using the currents, one can calculate the invariant amplitude for an electron-muon scattering presented in Fig.(\ref{figure1}). 
One can express the square of the invariant amplitude as
\begin{equation}\label{invariantamplitude}
    \left|\mathfrak{M}\right|^2=\frac{e^4}{-q^2(1-2\gamma q^2)}L_e^{\mu\nu}L^{muon}_{\mu\nu}\,,
\end{equation}
where the two fermions are scattered via the exchange of a gauge boson $A^\mu$ .
The tensors  $L_e^{\mu\nu}$ and $L^{muon}_{\mu\nu}$ can be expressed from Eqs.\eqref{currentsU} as follows 
\begin{equation}\label{tensorL}
    L_e^{\mu\nu}=\frac{1}{2}\sum_{spins}\left[j_{fi\,\mu}^{(1)} +j_{fi\,\mu}^{(2)} +j_{fi\,\mu}^{(3)}\right]\left[ j_{fi\,\nu}^{(1)} +j_{fi\,\nu}^{(2)} +j_{fi\,\nu}^{(3)}\right]^*\,,
\end{equation}
which, using Eq.\eqref{currentsU}, can be expressed as 
\begin{multline}\label{TensorAmpl}
    L_e^{\mu\nu}=\frac{1}{2}\sum_{spins}\left[-\bar{u}_f \tau^\mu u_i-2\gamma\bar{u}_f k_{f\,\rho}\tau^\rho k_i^\mu u_i+\gamma\bar{u}_f \tau^\mu k_{i\rho}k^{\rho}_i u_i\right]\\
    \times\left[-\bar{u}_f \tau^\nu u_i-2\gamma\bar{u}_f k_{f\,\rho}\tau^\rho k_i^\nu u_i+\gamma\bar{u}_f \tau^\nu k_{i\rho}k^{\rho}_i u_i\right]^*\,.
\end{multline}
The expression for $L^{muon}_{\mu\nu}$ is obtained analogously. 
Further expanding Eqs.\eqref{tensorL} and \eqref{TensorAmpl} and truncating to first order in $\gamma$,
one obtains
\begin{subequations}
\begin{multline}
     L_e^{\mu\nu}=\frac{1}{2}\sum_{spins}\left[(1-2\gamma k_\rho k^{\rho})\bar{u}(k') \tau^\mu u(k)\bar{u}(k) \tau^\nu u(k')\right.\\+2\gamma\bar{u}(k') \tau^\mu u(k)\bar{u}(k) k_{\rho}\tau^\rho k'^{\nu} u(k')\\\left.+2\gamma\bar{u}(k')k'_{\rho}\tau^\rho k^{\mu} u(k)\bar{u}(k) \tau^\nu u(k')\right]\,,
\end{multline}
 \begin{multline}
     L^{\text{muon}}_{\mu\nu}=\frac{1}{2}\sum_{spins}\left[(1-2\gamma p_\rho p^{\rho})\bar{u}(p') \tau_\mu u(p)\bar{u}(p) \tau_\nu u(p')\right.\\+2\gamma\bar{u}(p') \tau_\mu u(p)\bar{u}(p) p_{\rho}\tau^\rho p'_{\nu} u(p')\\\left.+2\gamma\bar{u}(p')p_{\rho}\tau^\rho p'_{\mu} u(p)\bar{u}(p) \tau_\nu u(p')\right]\,.
\end{multline}
\end{subequations}
Summing over the spins and taking the trace over the Dirac matrices, one obtains 
%
%
%
 \begin{subequations}\label{TracedTensors}
\begin{align}
\nonumber L_e^{\mu\nu}=2\left(1-2\gamma k\cdot k\right)&\left[k'^\mu k^\nu+k'^\nu k^\mu-\left(k'\cdot k-m_e^2\right)g^{\mu\nu}\right]\\+&8\gamma\left[\left(k\cdot k\right)k'^\mu k'^\nu+m_e^2 k'^{\mu}k^{\nu}\right]\\
\nonumber   L^{\text{muon}}_{\mu\nu}=2\left(1-2\gamma p\cdot p\right)&\left[p'_\mu p_\nu+p'_\nu p_\mu-\left(p'\cdot p-m_{\text{muon}}^2\right)g_{\mu\nu}\right]\\+&8\gamma\left[\left(p\cdot p\right)p'_\mu p'_\nu+ m_{\text{muon}}^2 p'_{\mu}p_{\nu}\right]
\end{align}
 \end{subequations}
Substituting the above Eqs.\eqref{TracedTensors} in Eq.\eqref{invariantamplitude}, one obtains the (squared) invariant amplitude including corrections up to linear order in $\gamma$
\begin{multline}
     \left\vert\mathfrak{M}\right\vert^2=  \frac{e^4}{\left(k-k'\right)^4}
     \left\{4\left[k'^\mu k^\nu+k'^\nu k^\mu-\left(k'\cdot k-m_e^2\right)g^{\mu\nu}\right]\left[p'_\mu p_\nu+p'_\nu p_\mu-\left(p'\cdot p-m_{\text{muon}}^2\right)g_{\mu\nu}\right]\right.\\\left.-8\gamma m_e^2\left[k'^\mu k^\nu+k'^\nu k^\mu-\left(k'\cdot k-m_e^2\right)g^{\mu\nu}\right]\left[p'_\mu p_\nu+p'_\nu p_\mu-\left(p'\cdot p-m_{\text{muon}}^2\right)g_{\mu\nu}\right]\right.\\\left.-8\gamma m_{\text{muon}}^2\left[k'^\mu k^\nu+k'^\nu k^\mu-\left(k'\cdot k-m_e^2\right)g^{\mu\nu}\right]\left[p'_\mu p_\nu+p'_\nu p_\mu-\left(p'\cdot p-m_{\text{muon}}^2\right)g_{\mu\nu}\right]\right.\\\left.+16\gamma m_e^2\left( k'^\mu k'^\nu+ k'^{\mu}k^{\nu}\right)\left[p'_\mu p_\nu+p'_\nu p_\mu-\left(p'\cdot p-m_{\text{muon}}^2\right)g_{\mu\nu}\right]\right.\\\left.+16\gamma m_{\text{muon}}^2 \left(p'_\mu p'_\nu+  p'_{\mu}p_{\nu}\right)\left[k'^\mu k^\nu+k'^\nu k^\mu-\left(k'\cdot k-m_e^2\right)g^{\mu\nu}\right]\right\}.
\end{multline}
Considering the above in the center of mass frame, applying the ultra-relativistic approximation to the leading order, {\it i.e.} the case in which the magnitude of the 3-momentum is much greater than the rest mass, that is $\vec{p}^{\,2}\gg m^2$ or $E^2\approx\vec{p}^{\,2}$.
Thus, mass terms in the leading order can be ignored.
However, the terms proportional to the mass in the corrections must be kept .
\begin{multline}
       \left|\mathfrak{M}\right|^2= \frac{8\,e^4}{\left(k-k'\right)^4}\left\{ \left(k'\cdot p'\right)\left(k\cdot p\right)+\left(k'\cdot p\right)\left(k\cdot p'\right) \right.\\-2\gamma\left(m_e^2+m_{\text{muon}}^2\right)\left[\left(k'\cdot p'\right)\left(k\cdot p\right)+\left(k'\cdot p\right)\left(k\cdot p'\right)
       \right]\\
    + 2 \gamma m_e^2 \left[\left(k'\cdot p'\right) \left(2 k'\cdot p + k\cdot p\right)
    + \left(k'\cdot k\right) \left(k\cdot p' - k'\cdot p\right)
       \right]
       \\
    \left. + 2 \gamma m_{\text{muon}}^2 \left[ \left(k'\cdot p'\right) \left(2 p'\cdot k + k\cdot p\right)
    + \left(k'\cdot k\right) \left(k\cdot p' - k'\cdot p\right) 
       \right]\right\}\,.
\end{multline}
Furthermore, in terms of the Mandelstam variables in the $s$ channel
\begin{subequations}
\begin{align}
    s&=\left(k+k'\right)^2=\left(p+p'\right)^2\approx2k\cdot k'\approx 2p\cdot p'\,,\\
    t&=\left(k-p\right)^2=\left(p'-k'\right)^2\approx -2k\cdot p\approx -2k'\cdot p'\,,\\
    u&=\left(k-p'\right)^2=\left(p-k'\right)\approx -2k\cdot p'\approx -2p\cdot k'\,,
\end{align}
\end{subequations}
the differential cross section reads
\begin{equation}
 \left.\frac{d\sigma}{d\Omega}\right\vert_{CM}=\frac{1}{64\pi^2\,s}\left\vert\mathfrak{M}\right\vert^2=\frac{2e^4}{64\pi^2\,s}\left[\frac{t^2+u^2}{s^2}+ \frac{1}{2}\gamma(m_e^2+m_{\text{muon}}^2)\frac{tu-u^2}{s^2} \right]\,,
 \end{equation}
which in terms of the energy and the scattering angle becomes
\begin{equation}
\left. \frac{d\sigma}{d\Omega} \right\vert_{CM} = \frac{\alpha^2}{4\,s} \left[ \frac{1}{2} \left(1 + \cos^2\theta\right) + \frac{1}{4} \gamma (m_e^2 + m_{\text{muon}}^2) \left(\cos\theta+\cos^2\theta\right)\right]\,.
 \end{equation}
Thus, one obtains the full cross-section by integrating over the full solid angle
 \begin{multline}
    \oint \left. \frac{d\sigma}{d\Omega} \right\vert_{CM}d\Omega \\
    = \int_0^{2\pi} \int_0^{\pi} \frac{\alpha^2}{4\,s} \left[ \frac{1}{2} \left(1 + \cos^2\theta\right) + \frac{1}{4} \gamma (m_e^2 + m_{\text{muon}}^2) \left(\cos\theta+\cos^2\theta\right)\right]\sin\theta d\theta d\phi\,,
 \end{multline}
 which gives
 \begin{equation}\label{totalcrosssection}
     \sigma\left(e^+e^-\rightarrow \mu^+\mu^-\right)=\frac{4\pi\alpha^2}{3\,s}\left[1+\frac{\gamma}{8}(m_e^2+m_{\text{muon}}^2)\right].
 \end{equation}
Finally, one can see that the total cross section is modified 
and the magnitude of the  correction depends only on the rest masses and fundamental constants.  
In fact if one assumes that QG effects are relevant at Planckian energies, 
equivalently if one assumes $\gamma_0=1$, the corrections for proton-proton scattering are about 
$\gamma m_p^2/4 \sim 10^{-38}$. 
The dependence of the corrections on the rest mass of the particles involved suggests that minimum length effects on scattering amplitudes may be measurable for electromagnetic scattering of heavier systems, such as scattering of heavy ions. The form of the RGUP corrections to the total cross section presented in Eq.\eqref{totalcrosssection} will hold true when considering scattering of heavy ions with total spin of the nucleus equal to  $1/2$. For example for the Xe-Xe scattering observed in the ATLAS experiment, the order of magnitude of the corrections is $\gamma m_{\text{Xe}}^2 /4\sim 10^{-34}$ \cite{Balek:2019nqk}.
 
\section{Conclusions}
In this work, expanding on \cite{Bosso2020:AoP}, we continued exploring the implications of a minimum length scale in quantum field theory. 
Employing the Ostrogradsky method, we derived  the RGUP modified fermion Lagrangian from its 
equations of motion, {\it i.e.} the RGUP modified Dirac equation.
We minimally coupled the resulting Lagrangian to an RGUP modified gauge field, constructed in such a way that
it preserves the usual $U(1)$ gauge symmetry. 
We then proceeded to examine the RGUP-QED Lagrangian Eq.\eqref{RGUPQED} and derived its Feynman rules.
A noticeable difference from the usual QED is that now up to five particle vertices are allowed.
In fact, all vertices contain two fermions and from one to three gauge bosons. 
We then computed the transition amplitudes and the fermionic current for the three particle vertices, which include the usual eletromagnetic scattering and RGUP corrections to it.
Then, we coupled two of these three particle vertices through a gauge boson and obtain the invariant amplitude of an electron muon scattering.
Considering that minimum length effects are expected to be relevant at very high energies, we considered the ultra-relativistic approximation.
Then, we calculate the differential and total cross section.\\
We observe interesting effects following from the existence of minimum length.
For example, in the ultra-relativistic limit, the only relevant correction  depends on the rest mass and  fundamental constants of nature. Therefore,  the minimum length effects on scattering amplitudes might be measurable using heavy ions scattering experiments like ATLAS.\\
The results presented in this work provide insight into testing minimum length effects in the laboratory. 
They also open the door for new calculations, such as the explorations of Higgs mechanism and symmetry breaking of RGUP-QFTs. This may lead to minimum length corrections to the mass of the Higgs field and by extension the masses of the gauge bosons and fermions.  Further research will  provide better limits on the scale at which QG effects are expected to be relevant.

\noindent
{\bf Acknowledgement}

 This work was supported by the Natural Sciences and Engineering Research Council of Canada, and University of Lethbridge. We thank M. Fridman and J. Stargen for providing feedback and valuable discussions.
 
\appendix 

 \section{\label{app:somutions}Dirac equation solutions}

In terms of the physical and the auxiliary momentum from the dispersion relation the Klein-Gordon (KG) equation reads as follows
\begin{equation}
     p^{\rho}p_{\rho}=  p_0^{\rho}p_{0\rho}(1+2\gamma p_0^{\sigma}p_{0\sigma}
    )=-(mc)^2 \,.
 \end{equation}
Since the KG equation remains unchanged in terms of the physical momentum, one can reasonably assume that the Dirac equation will have the same form 
  \begin{equation}\label{app:dirac}
   E\psi =\left(\vec\alpha \cdot \vec{p}+ \beta m\right)\psi\,,
 \end{equation}
where $E$ is the zeroth component and $\vec{p}$ is the spacial part of the physical momentum $p^\mu$. It can be proven that one has the following properties for $\vec\alpha$ and $\beta$
 %
 %
 %
 %
 %
 \begin{align}
    &\left\{\alpha_i,\,\alpha_j\right\}=\left\{\alpha_i,\,\beta\right\}=\left\{\beta,\,\beta\right\}=0\\
     &\alpha_i^2=\beta^2=1\,,
 \end{align}
 one can recall that these are the properties of Dirac matrices in fact one can recover a representation of the Dirac matrices form $\vec\alpha$ and $\beta$ as follows 
 \begin{equation}
   \tau^\mu\equiv \left(\beta,\beta\vec\alpha \right)\,,
 \end{equation}
which in terms of the Pauli matrices has the form presented in Eq.\eqref{gamma}.\\
%
%
%
%
%
In terms of linear operators the Dirac equation is can be written in the following form 
\begin{equation}
Hu=\left(\begin{array}{cc}
 m    & \vec\sigma\cdot\vec{p} \\
\vec\sigma\cdot\vec{p}     & -m
\end{array}\right)\left(\begin{array}{c}
     u_A  \\
     u_B
\end{array}\right)=E\left(\begin{array}{c}
     u_A  \\
     u_B
\end{array}\right)\,.\end{equation}
%
%
%
One can easily prove that Eqs.\eqref{freefieldsolutions}
\begin{subequations}
\begin{align}
    u^{(s)}\left(p\right)=N\begin{pmatrix}
\chi^{(s)}\\
\frac{\sigma \cdot p}{E+m}\chi^{(s)}
\end{pmatrix}\,, E>0\\
    u^{(s+2)}\left(p\right)=N\begin{pmatrix}
\frac{-\sigma \cdot p}{E+m}\chi^{(s+2)}\\
\chi^{(s+2)}
\end{pmatrix}\,, E<0\,.
\end{align}
\end{subequations}
are the respective solutions for a free Dirac field. Therefore  they can be used when calculating the transition amplitude. Note that $\vec{p}$ and $E$ are the physical momentum and energy, each of which can be expressed as a function of the auxiliary variables $\vec{p}_0$ and $E_0$.

\section{\label{app:lagrangian}Obtaining the Lagrangian}
To obtain the Lagrangian shown in Eq.\eqref{DirackLagrangian},the differential form of the Dirac equation is needed.
After the  substitution of  the expression for the physical momentum in terms of the auxiliary one
\begin{equation}
p_\mu=p_{0\,\mu}(1+\gamma p_{0\rho}p_0^\rho)\,,
\end{equation}
in Dirac equation  Eq.\eqref{EoM}.
One gets the following expression
\begin{equation}
    \left[\tau^{\mu}p_{0\,\mu}(1+\gamma p_{0\rho}p_0^\rho)-m\right]\psi=0\,.
\end{equation}
Because $p_0$ and $x_0$ are canonically conjugated to each other, one can write the differential form of the Dirac equation as
\begin{equation}\label{app:DiffDirac}
    \left[ i\tau^\mu\partial_\mu(1+\gamma\partial_\rho\partial^\rho) -m\right]\psi=0\,.
\end{equation}
Above, one can recognise  Euler-Lagrange equation for this theory.
Therefore, the Lagrangian can be recovered by  applying  the Ostrogradsky method \cite{deUrries:1998obu,Woodard:2015zca,Pons:1988tj} to obtain the Euler-Lagrange equations for theories with higher derivatives
\begin{equation}
    \frac{dL}{dq} -\frac{d}{dt}\frac{dL}{d\dot{q}}+\frac{d^2}{dt^2}\frac{dL}{d\ddot{q}}+\ldots+(-1)^n\frac{d^n}{dt^n}\frac{dL}{d (d^nq/dt^n)}=0\,,
\end{equation}
which in the case of fields is
\begin{equation}
    \frac{\partial\mathcal{L}}{\partial\phi}-  \partial_\mu  \frac{\partial\mathcal{L}}{\partial(\partial_\mu\phi)}+   \partial_{\mu_1} \partial_{\mu_2}\frac{\partial\mathcal{L}}{\partial(\partial_{\mu_1} \partial_{\mu_2}\phi)}+\ldots
    +(-1)^m\partial_{\mu_1}\ldots\partial_{\mu_m}\frac{\partial\mathcal{L}}{\partial(\partial_{\mu_1} \ldots\partial_{\mu_m}\phi)}=0\,.
\end{equation}
The first step is to assume a general form of the Lagrangian, where the order of derivatives is determined by the order of the equation of motion Eq.\eqref{app:DiffDirac}.
Moreover, an  assumption is made that different terms will have an arbitrary numerical coefficients multiplying every term
\begin{equation}\label{generalL}
     \mathcal{L}_{\psi}=\bar\psi\left[ iC_1\tau^\mu\partial_\mu(1+C_2\gamma\partial_\rho\partial^\rho) -C_3 m\right]\psi\,.
 \end{equation}
Next step is to prove that it has Eq.\eqref{app:DiffDirac} as an equation of motion.
Applying the Ostrogratsky method, one gets the following equations of motion for the field and its complex conjugated
\begin{align}
     &C_1i\tau^\mu\partial_\mu\psi +C_1C_2\gamma\partial_\rho\partial^\rho\psi -C_3 m\psi=0,\\
     &C_1i\tau^\mu\partial_\mu\bar\psi+C_1C_2\gamma\partial_\rho\partial^\rho\bar\psi -C_3 m\bar\psi=0\,.
\end{align}
The equations of motion obtained trough the Ostrogradsky method from Eq.\eqref{generalL} need to be identical to Eq.\eqref{app:DiffDirac}, which was obtained trough different means.  Therefore, the Lagrangian corresponding to the QFT spinor with minimum length will be of the form presented in Eq.\eqref{DirackLagrangian}.
This can be used to fix the value of the arbitrary coefficients. The result is an unique set of coefficients 
\begin{equation}
    C_1=C_2=C_3=1\,.
\end{equation}
Therefore, the Lagrangian corresponding to the QFT spinor with minimum length will be of the form presented in Eq.\eqref{DirackLagrangian}.

\end{document}